\documentclass[
aps,
floatfix,
letterpaper,
prl,
reprint]{revtex4-1}
\usepackage{CJK}
\usepackage{bbold}
\usepackage[utf8]{inputenc} 
\usepackage{braket}
\usepackage{ragged2e}
\usepackage{wrapfig}
\usepackage{pifont}
\usepackage{csvsimple}
\usepackage{setspace,dcolumn}
\usepackage[colorlinks=true, linkcolor=blue]{hyperref}
\usepackage{color}
\usepackage{graphicx,psfrag,epsfig}
\usepackage{subfigure}
\usepackage{amsmath,amsfonts,amssymb,amsthm,bm,upgreek}
\usepackage[mathscr]{eucal}
\usepackage[squaren]{SIunits}
\usepackage{url}
\usepackage{threeparttable}

\newcommand*{\B}[1]{\ifmmode\bm{#1}\else\textbf{#1}\fi}

\newcommand{\RNum}[1]{\uppercase\expandafter{\romannumeral #1\relax}}
\usepackage[colorinlistoftodos]{todonotes}

\begin{document}
\begin{CJK*}{GB}{} 
\title{Charge State Dynamics and Optically Detected Electron Spin Resonance Contrast of Shallow Nitrogen-Vacancy Centers in Diamond} 

\author{Zhiyang Yuan}
\affiliation{Department of Electrical Engineering, Princeton University, Princeton, New Jersey 08544, USA}
\author{Mattias Fitzpatrick}
\affiliation{Department of Electrical Engineering, Princeton University, Princeton, New Jersey 08544, USA}
\author{Lila V. H. Rodgers}
\affiliation{Department of Electrical Engineering, Princeton University, Princeton, New Jersey 08544, USA}
\author{Sorawis Sangtawesin}
\altaffiliation{Present address: School of Physics, Suranaree University of
Technology, Nakhon Ratchasima 30000, Thailand.}
\affiliation{Department of Electrical Engineering, Princeton University, Princeton, New Jersey 08544, USA}
\author{Srikanth Srinivasan}
\altaffiliation{Present address: IBM T.J. Watson Research Center, Yorktown Heights, NY 10598, USA.}
\affiliation{Department of Electrical Engineering, Princeton University, Princeton, New Jersey 08544, USA}
\author{Nathalie P. de Leon}
\email{npdeleon@princeton.edu}
\affiliation{Department of Electrical Engineering, Princeton University, Princeton, New Jersey 08544, USA}
\date{\today}

\begin{abstract}
Nitrogen-vacancy (NV) centers in diamond can be used for nanoscale sensing with atomic resolution and sensitivity; however, it has been observed that their properties degrade as they approach the diamond surface. Here we report that in addition to degraded spin coherence, NV centers within nanometers of the surface can also exhibit decreased fluorescence contrast for optically detected electron spin resonance (OD-ESR). We demonstrate that this decreased OD-ESR contrast arises from charge state dynamics of the NV center, and that it is strongly surface-dependent, indicating that surface engineering will be critical for nanoscale sensing applications based on color centers in diamond.

\end{abstract}
\maketitle 
\end{CJK*}

Nitrogen-vacancy (NV) centers in diamond are actively explored for a number of applications in quantum information processing and sensing because they exhibit long spin coherence times at room temperature, and their spin states can be optically initialized and read out with off-resonant excitation \cite{Hong2013, Sage2013, Maurer2012, Hensen2015, Humphreys2018,Lovchinsky2016}. In order to achieve strong interactions with materials and molecules that are external to the diamond, NV centers must be placed close to the diamond surface. It has been well established that the diamond surface can host contaminants, magnetic defects, and electronic defects that give rise to noise, leading to short spin coherence times \cite{Sangtawesin2019a, Myers2014, DeOliveira2017, Grotz2011, Ohno2012, Naydenov2010, Ofori-Okai2012, Kim2015, Rosskopf2014, Romach2015}, and recent work has shown that careful preparation of the diamond surface can mitigate this noise, leading to extended spin coherence times \cite{Sangtawesin2019a,DeOliveira2017,Lovchinsky2016}. Most NV center applications are based on the negative charge state of NV centers, NV$^-$, and it is also known that NV centers can exist in the neutral charge state, NV$^0$. It has been recently shown that both the steady-state charge populations under illumination and the equilibrium charge state in the dark of shallow NV centers can vary significantly from bulk properties  \cite{Bluvstein2019, Stacey2019, Dhomkar2018, Kageura2017}. Previous work has shown that charge state initialization can influence spin readout \cite{Hopper2020}, however a detailed understanding of how these charge state dynamics can affect schemes for nanoscale sensing and its dependence on sample surface has not yet been established.

Here we demonstrate that the diamond surface can strongly affect both the steady-state charge state distribution and the charge state dynamics of NV centers within nanometers of the diamond surface, and that these charge state properties can significantly degrade optically detected electron spin resonance (OD-ESR) contrast. We focus on two diamond samples (samples A and F) that contain shallow NV centers introduced by ion implantation, which exhibit distinctly different OD-ESR contrast and charge state behavior. We find that sample A exhibits higher contrast, greater charge state stability in the dark, lower NV$^0$ population under green illumination, and slower ionization and recombination dynamics than sample F. More specifically, we find that sample F exhibits charge state conversion rates that are comparable to the internal spin-dependent dynamics of the NV center, thus leading to decreased OD-ESR contrast. Finally, we measure time-resolved, spin-dependent photoluminescence (PL) and compare to a coupled rate equation model that incorporates charge conversion rates to quantitatively understand the impact of charge dynamics on OD-ESR contrast.

\begin{figure}[tb]
	\centering
	\includegraphics[width=8.6 cm]{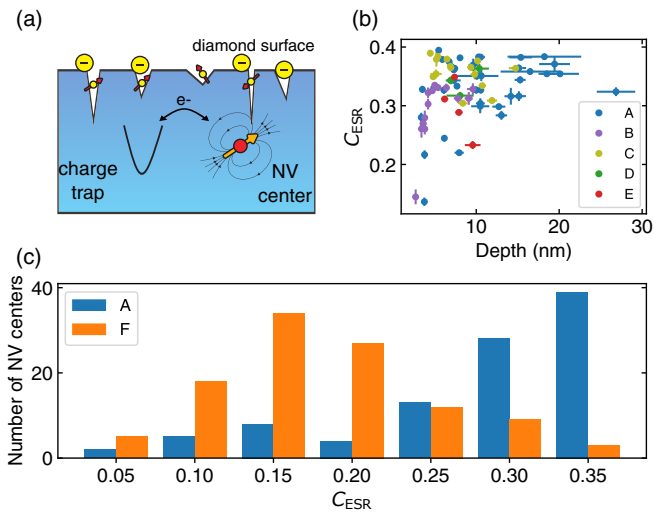}
	\caption{Variation in OD-ESR contrast for shallow NV centers. (a) Cartoon depicting the interaction between shallow NV centers and defects at the diamond surface. Defects can act as charge traps, leading to changes in ionization and recombination kinetics for nearby NV centers. (b) OD-ESR contrast, $C_{\text{ESR}}$, as a function of depth for NV centers across five diamond samples. A large spread and an average decrease in contrast is apparent within around 10 nm of the surface. (c) Histograms of $C_{\text{ESR}}$ from the two selected samples, A (blue bars) and F (orange bars). 99 NV centers in sample A and 108 NV centers in sample F were measured under the same experimental conditions including microwave pulse power and green laser power. An external magnetic field of approximately 27 G was aligned to the NV axis.}
	\label{fig:fig1}
\end{figure}

A cartoon schematic of the interactions between shallow NV centers and charge traps is depicted in Fig.~\ref{fig:fig1}(a). The surface can host defects that act as charge traps, and tunneling from NV centers to these charge traps gives rise to charge state instability, leading to a decrease in OD-ESR contrast of shallow NV centers \cite{Bluvstein2019}.

Following Ref.~\cite{Hopper2018}, we define OD-ESR contrast as 

\begin{equation}
    C_{\text{ESR}} = \frac{\alpha_0 - \alpha_{\pm1}}{\alpha_{0}},
\end{equation}

\noindent where $\alpha_i$ is the average number of photons collected in the readout pulse when the NV$^-$ spin state is initialized to the $m_{s} = i$ state. Figure~\ref{fig:fig1}(b) shows $C_{\text{ESR}}$ plotted for many NV centers as a function of depth across five samples, where depth is measured by detecting the proton NMR signal arising from the microscope immersion oil \cite{Pham2016}. NV centers within \unit{10}{\nano\meter} of the surface display a wider distribution in $C_{\text{ESR}}$.
 
In order to investigate the origin of the lower OD-ESR contrast, we identify two samples that have undergone nominally similar surface processing (see Supplemental Material Sec.~\RNum{1} \cite{SI}), but exhibit markedly different distributions in $C_{\text{ESR}}$, samples A and F [Fig.~\ref{fig:fig1}(c)]. Most NV centers in sample F exhibit $C_{\text{ESR}}$ below 0.3, as well as short coherence times ($T_2 < \unit{4}{\micro\second}$), precluding many NV characterization measurements, such as single-shot charge state readout \cite{Aslam2013} and using proton NMR to measure depth. We use this comparison to quantify the impact of charge state conversion on $C_{\text{ESR}}$.

Both samples were prepared using nitrogen ion implantation followed by thermal annealing. Sample A was implanted with a nitrogen ion energy of \unit{3}{\kilo\electronvolt} while sample F was implanted with an energy of \unit{1.5}{\kilo\electronvolt}. Despite undergoing nominally similar surface processing, sample F contains persistent boron contamination comprising approximately 4$\%$ of a surface monolayer as measured by X-ray photoelectron spectroscopy \cite{SI}, likely arising from contamination in the furnace during vacuum thermal annealing. We note that although sample F is expected to have a shallower distribution of NV centers because of its lower implantation energy, the distribution of $C_{\text{ESR}}$ is much lower than another sample that was prepared with the same ion implantation energy that does not exhibit surface contamination, sample B \cite{SI}.


\begin{figure}[ht]
	\centering
	\includegraphics[width=8.6 cm]{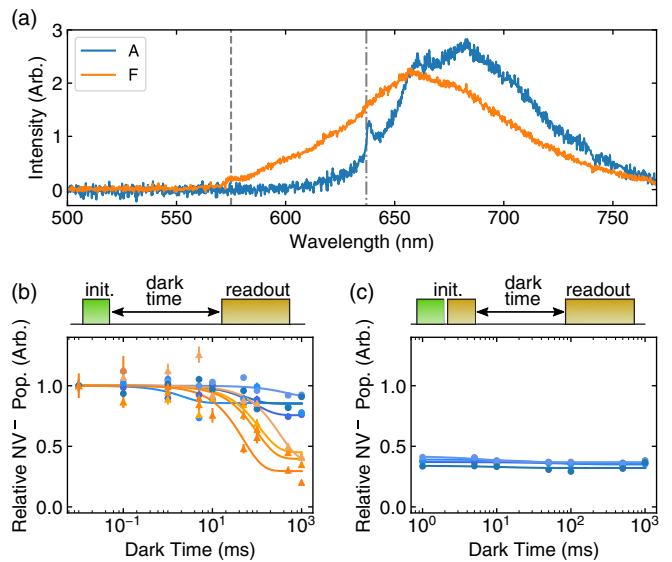}
	\caption{Steady-state charge distributions. (a) PL spectrum under green illumination for representative NV centers in sample A (blue trace) and sample F (orange trace). The higher level of PL between 575 nm (dashed line) and 637 nm (dash-dotted line) in sample F indicates a higher steady-state population in NV$^0$. (b) Charge state conversion in the dark. NV centers are initialized into the negative charge state by a green pulse, and the charge state is read out using a charge-state-selective orange pulse with variable delay time. The PL is normalized to the value for the shortest dark time interval. Each NV center in sample F (orange triangles) exhibits a decay in the relative NV$^-$ population with a time constant in the range of 11 - 300 ms, while NV centers in sample A (blue circles) exhibit much less decay out to 1 s. Solid lines are fits to the data. (c) Stability of NV$^0$ in the dark in sample A. NV centers are initialized into NV$^0$ using a green initialization pulse followed by an orange pulse, and then read out with another orange pulse after a variable delay time. The PL is normalized to the value after the green pulse. No change in the population is observed out to 1 s, indicating that recombination kinetics are also slow in sample A.}
	\label{fig:fig2}
\end{figure}

To understand the origin of the lower OD-ESR contrast in sample F, we examine the steady-state charge distributions in both samples. First we measure the PL spectrum under green illumination. We observe a significantly higher fraction of emission between the zero phonon line (ZPL) of NV$^0$ (575 nm) and the ZPL of NV$^-$ (637 nm) in sample F, as well as a peak at 575 nm [Fig.~\ref{fig:fig2}(a)], indicating a higher steady-state population of NV$^0$ \cite{Alsid2019a}. This higher NV$^0$ steady-state population contributes to background fluorescence, decreasing the OD-ESR contrast proportionally. 

Recent work has shown that shallow NV centers can exhibit spontaneous conversion to NV$^0$ without illumination \cite{Bluvstein2019}. Here, we perform a similar measurement on several NV centers in both samples. First, the NV center charge state is initialized using a green pulse of sufficient length (\unit{5}{\milli\second} at \unit{510}{\micro\watt}) to achieve a steady-state distribution. Then, after a variable time interval in the dark, we measure PL during an orange (\unit{590}{\nano\meter}) pulse, which preferentially excites NV$^-$, allowing for charge state readout. Normalizing to the PL at the shortest dark time interval, we observe that NV centers in sample F exhibit a decay in the NV$^-$ population over timescales between $\unit{11}{\milli\second}$ and $\unit{300}{\milli\second}$ to less than half of the initial value, while a decay of less than 25$\%$ is observed in sample A for dark times up to 1 second [Fig.~\ref{fig:fig2}(b)]. In sample A, we also perform the inverse measurement by initializing NV centers into NV$^0$ using an orange pulse (\unit{5}{\milli\second} at \unit{18}{\micro\watt}, after a green pulse as Fig.~\ref{fig:fig2}(b) for consistency \cite{Dhomkar2018}) to look for evidence of spontaneous conversion to NV$^-$. We observe no change in the NV$^-$ population out to 1 second [Fig.~\ref{fig:fig2}(c)], indicating that spontaneous charge conversion is very slow in this sample, regardless of the initial charge state population. This is consistent with previous hypotheses that spontaneous charge conversion is mediated by the availability of nearby electron traps at or near the surface [Fig.~\ref{fig:fig1}(a)] that strongly modify charge state kinetics \cite{Bluvstein2019}.

Now we turn to the charge state conversion dynamics under illumination for both samples. The rate of change of the charge populations can be expressed in the following simplified model \cite{Aslam2013}:

\begin{equation}
\label{equ:charge_rate_eq}
\left\{ 
\begin{aligned}
\frac{d{\rho}_-}{dt} =& -r_{\text{ion}} \rho_- + r_{\text{rec}} \rho_0\\
\frac{d{\rho}_0}{dt} =& r_{\text{ion}} \rho_- - r_{\text{rec}} \rho_0
\end{aligned}
\right. ,
\end{equation}
where $\rho_-$ is the NV$^-$ population, $\rho_0$ is the  NV$^0$ population, and $r_{\text{ion}}$ and $r_{\text{rec}}$ are ionization and recombination rates, respectively. Since the total population is conserved ($\rho_- + \rho_0 = 1$), we can express the time dependence of the population in terms of a total charge conversion rate, $\rho_{-,0} (t) \propto e^{-r_{\text{tot}} t}$, where $r_{\text{tot}} = r_{\text{ion}} +r_{\text{rec}}$. 

\begin{figure}[tb]
	\centering
	\includegraphics[width=8.6 cm]{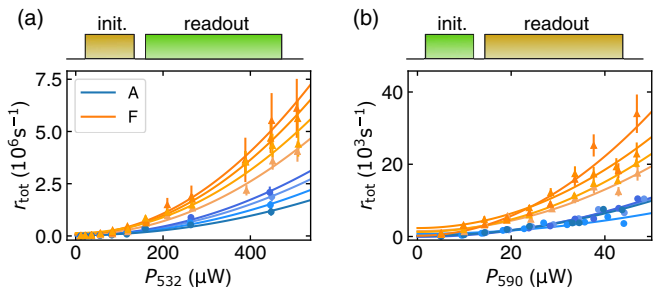}
	\caption{Charge state dynamics for NV centers in sample A (blue circles) and sample F (orange triangles). In both plots, points are experimental data and lines are fit using a quadratic function. (a) Power dependence of the charge state conversion rates under green illumination. (b) Power dependence of the charge state conversion rates under orange illumination. Laser powers are measured at the back of the objective. Under both illumination conditions, the charge state conversion rates are faster for each NV center in sample F compared to those in sample A.}
	\label{fig:fig3}
\end{figure}

We measure charge conversion rates under green illumination by using a fixed orange pulse to initialize the NV centers primarily into NV$^0$. Then under green illumination, the population shifts to NV$^-$, and we fit the time-resolved PL to an exponential to extract $r_{\text{tot}}$. Similarly, we measure conversion rates under orange illumination by first initializing with a green pulse and measuring the overall ionization under orange, extracting $r_{\text{tot}}$ in the same manner. Both data sets are plotted in  Fig.~\ref{fig:fig3} for a set of NV centers in samples A and F. We observe a clear quadratic power dependence for NV centers in both samples, consistent with both ionization and recombination being two-photon processes \cite{Aslam2013} (see Supplemental Material Sec.~\RNum{6} \cite{SI} for a simplified derivation). NV centers in sample F have higher charge conversion rates under both green and orange illumination. Importantly, when the green power is comparable to the typical NV saturation power in our setup ($\approx\unit{300}{\micro\watt}$), the charge state conversion rate can be faster than \unit{1\times10^6}{\second^{-1}}, comparable to the spin polarization rate for NV$^-$, suggesting that fast ionization and recombination processes can occur during the spin initialization and readout times. Since these processes are not spin conserving \cite{Roberts2019}, this fast charge conversion can lead to degradation of the OD-ESR contrast.

\begin{figure}[!t]
	\centering
	\includegraphics[width=8.6 cm]{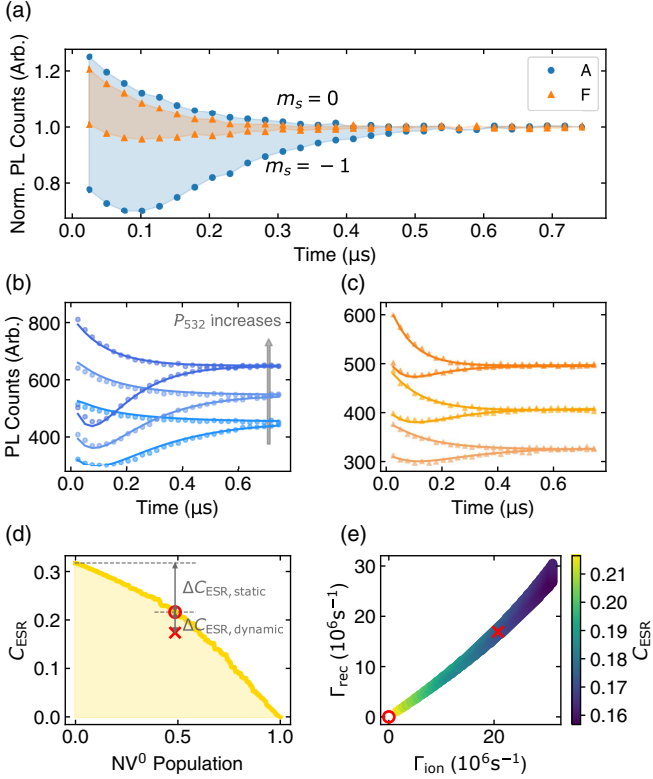}
	\caption{Time-resolved PL traces and fitting to rate equation model. (a) Comparison of PL traces for representative NV centers from the two samples, A (blue) and F (orange), under \unit{664}{\micro\watt} of green illumination. PL counts are normalized to steady-state values for each trace. The shaded area is proportional to the OD-ESR contrast. (b,c) Power dependence of NV PL traces from sample A (b) and sample F (c) after initialization into the NV$^-$ $m_s = 0$ state and $m_s = -1$ state. Each trace represents a different green illumination power (from bottom to top: \unit{330}{\micro\watt}, \unit{450}{\micro\watt}, \unit{660}{\micro\watt}). Solid lines indicate fits to the data using the rate equation model. (d) Simulated contrast as a function of NV$^0$ population using extracted parameters from the fit to the same NV as in (c), where the upper bound (dark yellow points) represents the highest attainable contrast for a particular NV$^0$ population and the shaded region represents lower $C_{\text{ESR}}$ caused by faster charge state conversion. The red cross indicates the extracted $C_{\text{ESR}}$ and NV$^0$ population for this NV center, and the highest attainable contrast for this NV$^0$ population is indicated by the red circle. The total decrease in contrast compared to $C_{\text{ESR}}$ with no charge dynamics is parameterized by the change due to the finite NV$^0$ population of 0.486, $\Delta C_{\text{ESR,static}}$, and the change due to charge conversion rates, $\Delta C_{\text{ESR,dynamic}}$. (e) $C_{\text{ESR}}$ as a function of $\Gamma_{\text{ion}}$ and $\Gamma_{\text{rec}}$ for a fixed NV$^0$ population ($\pm 1\%$). The red cross indicates the extracted $\Gamma_{\text{ion}}$ and $\Gamma_{\text{rec}}$ and the red circle indicates the point of highest $C_{\text{ESR}}$.}
	\label{fig:fig4}
\end{figure}

To quantify how charge state conversion rates affect OD-ESR contrast for NV centers under distinctly different surface environments, we measure the time-resolved PL during the green readout pulse. Example PL traces with initial NV$^-$ spin states $m_s=0$ (upper traces) and $m_s=-1$ (lower traces) for NV centers in both samples are shown in Fig.~\ref{fig:fig4}(a). The area between the two sets of curves is proportional to $C_{\text{ESR}}$, and in a typical sensing experiment, the readout conditions would be optimized to maximize this area. We observe that the total area between the curves for sample A is $\sim$3.6 times the area between the curves for sample F. We fit the data using a rate equation model for NV$^-$ spin dynamics \cite{Hacquebard2018, Robledo2011, Roberts2019, Manson2006}, modifying the model to additionally incorporate NV$^0$ states and charge state conversion rates $\Gamma_{\text{ion}}$ and $\Gamma_{\text{rec}}$ (see a complete description of the rate equation model in Supplemental Material Sec.~\RNum{3} \cite{SI}). In order to constrain the model, we measure PL traces for both $m_s=0$ and $m_s={-1}$ initial states for eight different laser powers and fit all data sets simultaneously to a single model. We constrain the fit to incorporate the same intrinsic NV parameters (excited state lifetime, singlet state lifetime, etc.) across all powers, while allowing the photo-induced transition rates to be free parameters. Representative PL traces for a subset of these powers for an NV center from sample A  are shown in Fig.~\ref{fig:fig4}(b), and corresponding datasets from sample F are shown in Fig.~\ref{fig:fig4}(c). Our fitted intrinsic NV parameters are consistent with previously reported measurements \cite{Robledo2011, Roberts2019}. We also find that extracted excited state and singlet lifetimes are consistent across NV centers. A summary of the extracted parameters for four NV centers in each sample can be found in Supplemental Material Sec.~\RNum{4} \cite{SI}.

From the data it is clear that there are two different contributions of charge state dynamics to decreased $C_{\text{ESR}}$: the increased background due to the steady-state NV$^0$ population, and charge conversion rates interfering with spin dynamics. Armed with a model that quantitatively accounts for both spin and charge dynamics of the NV center, we use this model to disentangle the relative contributions to $C_{\text{ESR}}$ of these two effects. Using the fitted NV parameters, we vary $\Gamma_{\text{ion}}$ and $\Gamma_{\text{rec}}$ to simulate values of $C_{\text{ESR}}$ and the NV$^0$ population. Using the extracted parameters for the NV center in sample F shown in Fig.~\ref{fig:fig4}(c), we plot the maximum $C_{\text{ESR}}$ as a function of steady-state NV$^0$ population [Fig.~\ref{fig:fig4}(d)]. As expected, $C_{\text{ESR}}$ decreases monotonically as the NV$^0$ population increases, and we define $\Delta C_{\text{ESR,static}}$ to be the decrease in $C_{\text{ESR}}$ due to the finite NV$^0$ population. 

Separately, as the overall $\Gamma_{\text{ion}}$ and $\Gamma_{\text{rec}}$ increase in magnitude, the charge conversion dynamics begin to compete with the spin dynamics, leading to an additional decrease in $C_{\text{ESR}}$. We define this additional decrease at a given NV$^0$ population as $\Delta C_{\text{ESR,dynamic}}$. We plot a contour of the simulated $C_{\text{ESR}}$ as a function of the charge state conversion rates at a fixed NV$^0$ population [Fig.~\ref{fig:fig4}(e)], and we also observe a monotonic decrease of $C_{\text{ESR}}$ with increasing $\Gamma_{\text{ion}}$ and $\Gamma_{\text{rec}}$. 

For this particular NV center in sample F, where $C_{\text{ESR}} = 0.174$, we would expect an improvement to $C_{\text{ESR}} = 0.317$ in the absence of any charge conversion, where $\Delta C_{\text{ESR,static}} = 0.100$ and $\Delta C_{\text{ESR,dynamic}} = 0.042$. By comparison, for the NV center from sample A shown in Fig.~\ref{fig:fig4}(b) with $C_{\text{ESR}} = 0.420$, the absence of charge conversion would yield a much smaller improvement in $C_{\text{ESR}}$ to 0.453, $\Delta C_{\text{ESR,static}} = 0.018$ and $\Delta C_{\text{ESR,dynamic}} = 0.015$.

We have shown that samples with different surface conditions can exhibit drastically different charge state dynamics for shallow NV centers, and that charge state stability can influence OD-ESR contrast in two ways: first, a high steady-state population in NV$^0$ will increase the PL background and second, when the charge state conversion time is comparable to the NV spin polarization and readout time, ionization results in a loss of spin information and initialization, and consequently a decrease of $C_{\text{ESR}}$. These effects can be drastically different for different diamond samples, as well as among NV centers in a given sample. These results point to the importance of surface preparation and engineering in utilizing NV centers for nanoscale sensing and magnetometry. On-going and future work includes establishing microscopic mechanisms for charge state instabilities, such as surface trap states and contaminants, as well as exploring optimal surface terminations \cite{Kaviani2014,Hauf2011,Stacey2015,Cui2013} for stabilizing the negative charge state of shallow NV centers.

\vskip 0.2in
We thank James J. Allred for help with sample processing, Trisha Madhavan for help with part of the data acquisition, Haimei Zhang for help implementing the charge state readout optical setup, and Nan Yao, Yao-Wen Yeh, and John Schreiber at the Princeton Imaging and Analysis Center for help with diamond surface characterization. We also thank Dolev Bluvstein, Bo Dwyer, and Shimon Kolkowitz for helpful discussions.
This material is based upon work supported by the U.S. Department of Energy, Office of Science, Office of Basic Energy Sciences, under Award Number DE-SC0018978. This work was also supported by the NSF under the CAREER program (Grant No. DMR-1752047), and was partially supported by the DARPA DRINQS program (Agreement No. D18AC00015).
M. F. was supported by an appointment to the Intelligence Community Postdoctoral Research Fellowship Program by Oak Ridge Institute for Science and Education (ORISE) through an interagency agreement between the U.S. Department of Energy and the Office of the Director of National Intelligence (ODNI). L. V. H. R. acknowledges support from the Department of Defense through the National Defense Science and Engineering Graduate Fellowship Program. 
\bibliographystyle{apsrev4-1}
\bibliography{chargeState.bib}
\end{document}


\begin{CJK*}{GB}{} 
\title{Supplemental Material for \\``Charge State Dynamics and Electron Spin Resonance Contrast of Shallow Nitrogen-Vacancy Centers in Diamond''} 

\author{Zhiyang Yuan}
\affiliation{Department of Electrical Engineering, Princeton University, Princeton, New Jersey 08544, USA}
\author{Mattias Fitzpatrick}
\affiliation{Department of Electrical Engineering, Princeton University, Princeton, New Jersey 08544, USA}
\author{Lila V. H. Rodgers}
\affiliation{Department of Electrical Engineering, Princeton University, Princeton, New Jersey 08544, USA}
\author{Sorawis Sangtawesin}
\affiliation{Department of Electrical Engineering, Princeton University, Princeton, New Jersey 08544, USA}
\author{Srikanth Srinivasan}
\affiliation{Department of Electrical Engineering, Princeton University, Princeton, New Jersey 08544, USA}
\author{Nathalie P. de Leon}
\email{npdeleon@princeton.edu}
\affiliation{Department of Electrical Engineering, Princeton University, Princeton, New Jersey 08544, USA}
\maketitle 
\end{CJK*}

\section{\romannumeral 1. Diamond sample preparation} 
\label{sub:samples}

Most diamond samples in this work are prepared following the procedure outlined in Ref.~\cite{Sangtawesin2019a}. For samples A, C, D and F, we started with electronic grade diamonds from Element Six which were scaife polished to a RMS roughness of less than \unit{1}{\nano\meter}. We then performed inductively-coupled plasma reactive ion etching (ICP-RIE) followed by high temperature annealing at \unit{1200}{\celsius} in vacuum to remove surface and subsurface damage. After annealing, samples were cleaned in a refluxing mixture of 1:1:1 concentrated sulfuric, nitric, and perchloric acids (triacid clean) to remove surface amorphous carbon formed during annealing. Samples B and E did not go through this polish and pre-etch process and so their surfaces were `as-grown' before implantation. Next, NV centers were formed by nitrogen ion implantation followed by annealing at \unit{800}{\celsius} in vacuum. Further details about implantation parameters for different samples are listed in Table~\ref{tab:tab1_Samples}. Afterwards, samples were triacid cleaned again. This was the last step performed for sample B. For samples A, C, D, E and F we subsequently annealed them in an oxygen atmosphere at temperatures around $\unit{440}{\degreecelsius}- \unit{460}{\degreecelsius}$ to create well-ordered oxygen-terminated diamond surfaces. After oxygen annealing, these samples were cleaned in Piranha solution (1:2 mixture of hydrogen peroxide and concentrated sulfuric acid).

\begin{table}[ht]
	\centering
	\begin{ruledtabular}
	\begin{tabular}{cccc}
	Sample & Surface condition & Implantation dose (cm$^{-2}$) & Implantation energy (keV)\\
	\colrule
	A & Polished + Pre-etched, $^{12}$C enriched & $5\times10^{8}$ & 3\\
	B & As-grown, $^{12}$C enriched & $1\times10^{9}$ & 1.5\\
	C & Polished + Pre-etched & $5\times10^{8}$ & 3\\
	D & Polished + Pre-etched & $1\times10^{9}$ & 3\\
	E & As-grown, $^{12}$C enriched & $3\times10^{9}$ & 3\\
    F & Polished + Pre-etched & $1\times10^{9}$ & 1.5\\
	\end{tabular}
	\end{ruledtabular}
	\caption{Sample ion implantation details.}
	\label{tab:tab1_Samples}
\end{table}

\begin{figure}[ht]
	\centering
	\includegraphics[width=8.6 cm]{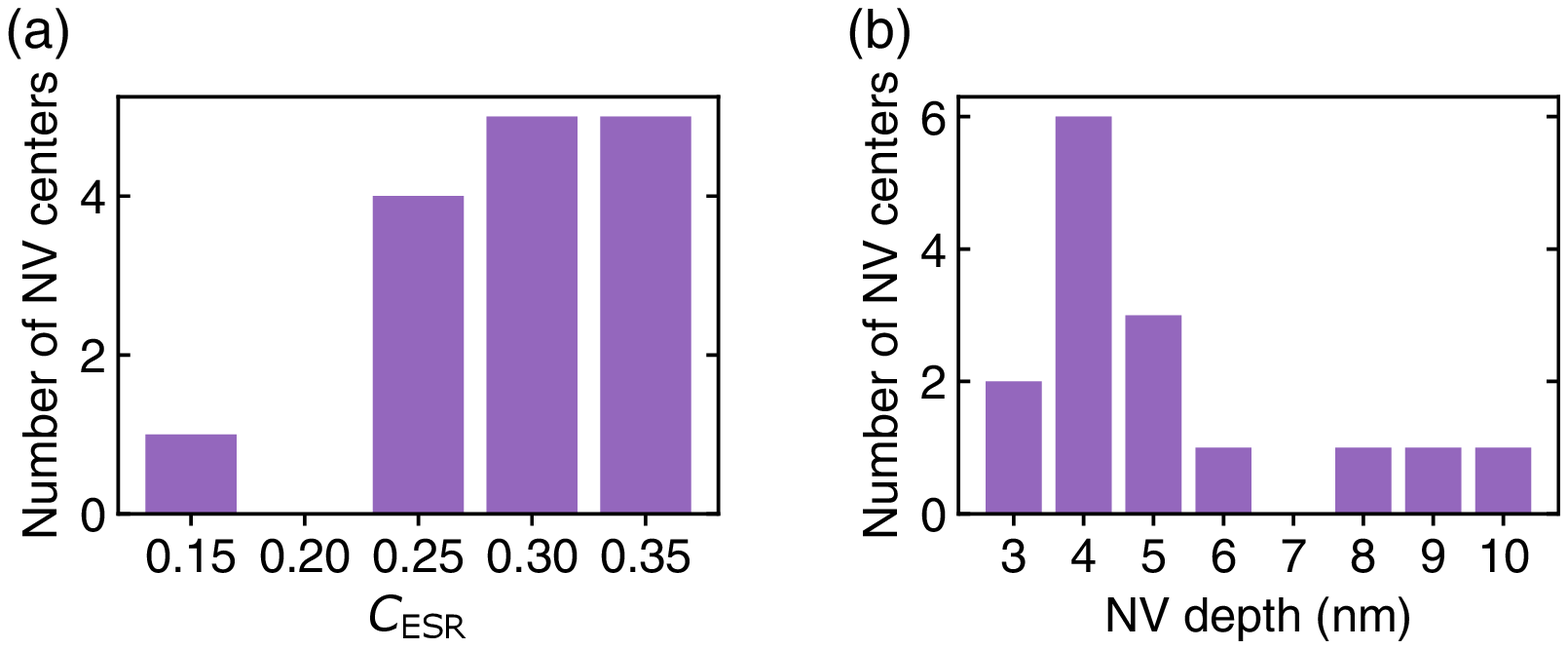}
	\caption{Characteristics of NV centers in sample B, which was implanted at the same energy as sample F but has an as-grown surface. (a) Histogram of OD-ESR contrast, $C_{\text{ESR}}$. (b) Histogram of measured NV depths.}
	\label{fig:s_fig sample B}
\end{figure}

Due to poor OD-ESR contrast in sample F, we are unable to use proton NMR to measure the depths of individual NV centers. Instead we provide a comparison sample that was implanted at the same energy, sample B. Histograms of the contrast and depth distributions for sample B are presented in Fig.~\ref{fig:s_fig sample B}. Although sample B was implanted with the same ion implantation energy as sample F, we observe significantly higher OD-ESR contrast for NV centers in this sample, which suggests the lower contrast in sample F mainly results from the diamond surface condition.


\begin{figure}[ht]
	\centering
	\includegraphics[width=8.6 cm]{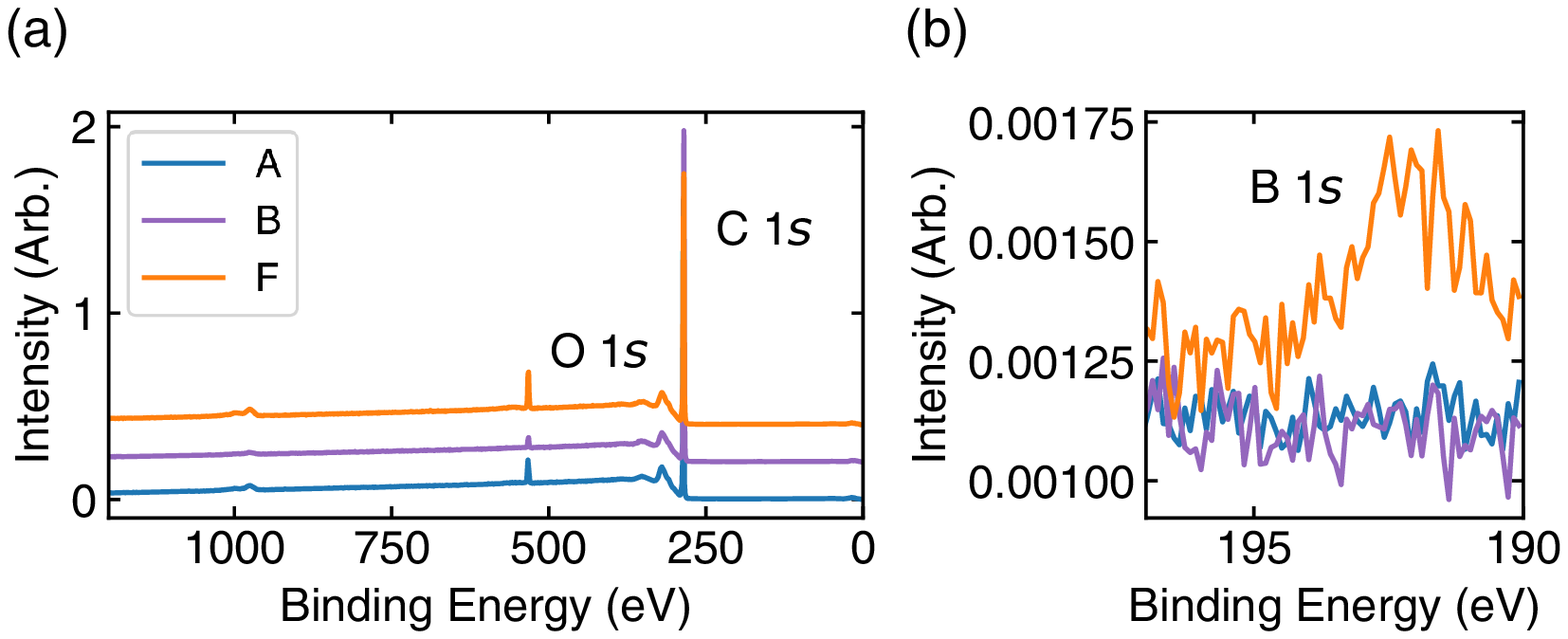}
	\caption{XPS characterization of surfaces for samples A, B and F. (a) XPS survey scans showing dominant carbon and oxygen peaks, shifted vertically for clarity. Samples A and F were annealed in an oxygen atmospher and show higher oxygen 1$s$ peaks. (b) High-resolution XPS boron 1$s$ spectrum showing a small but clear peak in sample F. Intensity is normalized by setting the height of carbon 1$s$ peak to 1.}
	\label{fig:s_fig_xps}
\end{figure}

X-ray photoelectron spectroscopy (XPS) characterization of the diamond surfaces was performed with a Thermo Fisher K-Alpha spectrometer at the Imaging and Analysis Center (IAC) at Princeton University. XPS spectra for samples A, B and F are shown in Fig.~\ref{fig:s_fig_xps}. A peak associated with boron 1$s$ is evident in sample F suggesting boron contamination on the surface, which has an atomic percentage of 0.31\%. We estimate that this atomic percentage corresponds to approximately 4\% of a surface monolayer of boron using the method discussed in Ref.~\cite{Sangtawesin2019a}.


\section{\romannumeral 2. Experimental Setup}\label{subsec:experimental_Setup}
NV photoluminescence (PL) measurements are performed in a home-built confocal microscope setup. A Nikon Plan Fluor $100\times$, $N.A.= 1.30$, oil immersion objective is used for focusing the excitation laser and collecting the PL. For the green illumination, we use a \unit{532}{\nano\metre} optically pumped solid-state laser (Coherent Sapphire LP 532-300), and for the orange illumination we use a NKT SuperK laser (repetition rate 78 MHz, pulse width 5 ps) with two bandpass filters with transmission wavelength around \unit{590}{\nano\metre} (Thorlabs FB590-10 and Semrock FF01-589/18-25). Both lasers are modulated by acousto-optic modulators (AOMs) (Isomet 1205C-1) and the beam is scanned with X-Y galvo mirrors (Thorlabs GVS012). A dichroic mirror (Semrock BLP01-647R-25) is used to separate the excitation and collection pathways, and the PL is measured using a fiber-coupled avalanche photodiode (Excelitas SPCM-AQRH-44-FC). 

For the PL spectrum measurements shown in the main text Fig.~2, we use a Princeton Instruments Monochromator (Acton SP-2300i) with a CCD camera (Pixis 100). Each spectrum is obtained by first acquiring the background signal of a non-NV spot close to the target NV and subtracting this background spectrum from the acquired NV spectrum. This technique allows us to remove background fluorescence and Raman lines of the driving laser.

All the pulse sequences used in the experiments are programmed by a PulseBlaster (SpinCore ESR-PRO500 with a timing resolution of \unit{2}{\nano\second}). For time-resolved measurements of NV PL, the avalanche photodiode signal is sent to a PicoHarp (PicoQuant PicoHarp 300 with a highest resolution of \unit{4}{\pico\second}).

\section{\romannumeral 3. Time-Resolved Measurements: Rate Equation Model and Data Fitting} 
\label{sub:time_resolved_measurement}
In our time-resolved measurements of NV PL, a green pulse is used to initialize the NV$^-$ spin state to $m_s=0$. Then we either leave the NV in $m_s=0$ or apply a calibrated microwave $\pi$ pulse to initialize the NV into $m_s=-1$. A subsequent green pulse is used to measure the PL which is read out by a PicoHarp to achieve high time resolution. We vary the power of both initialization and readout green pulses together to get the power dependence of the time-resolved PL. It is worth noting that this potentially changes the initial state that our model uses to predict the time-resolved PL. We address this in our modeling.

\begin{figure}[tb]
	\begin{center}
		\includegraphics[width=8.6 cm]{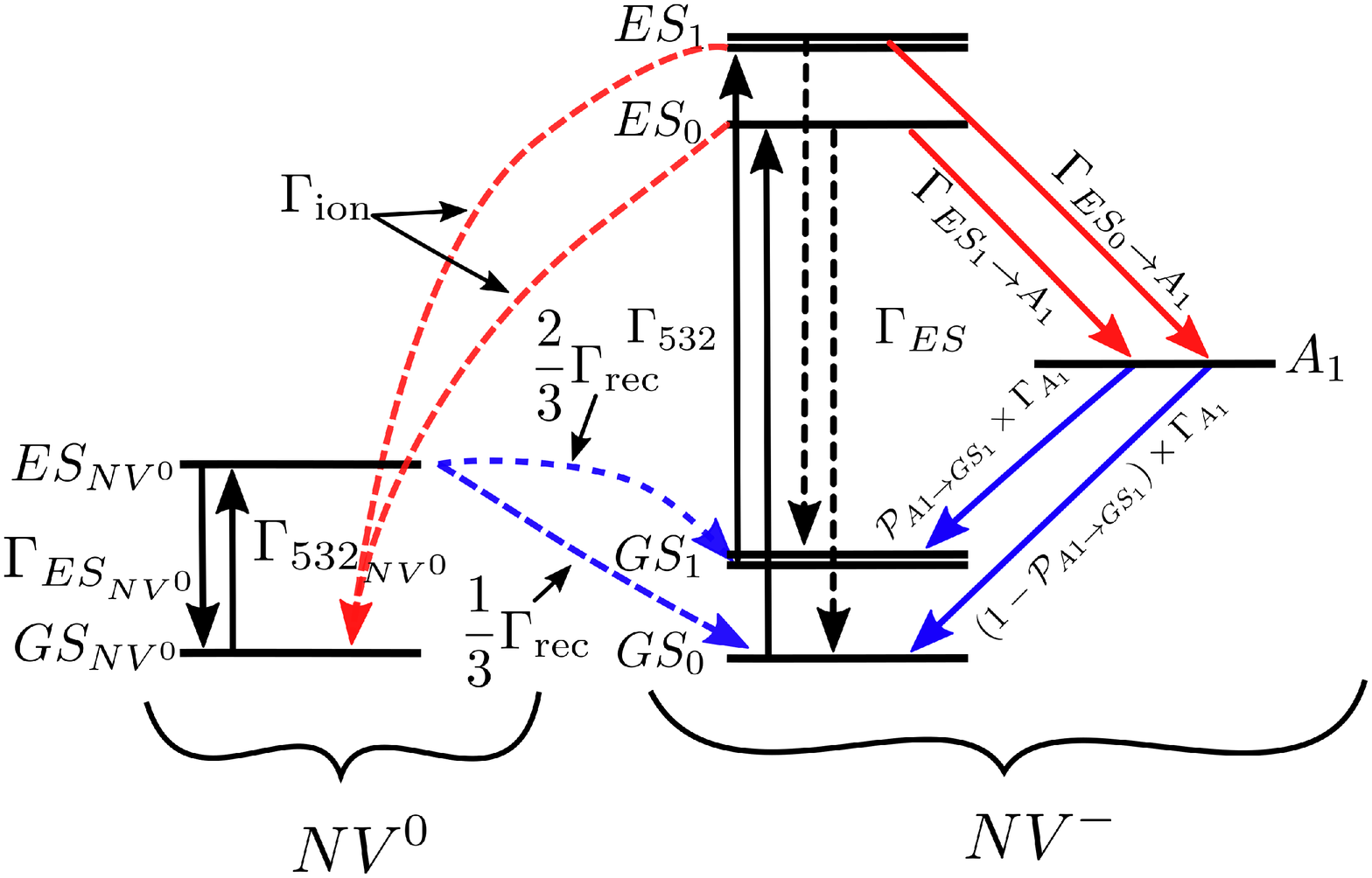}
	\end{center}
	\vspace{-0.6cm}
	\caption{ NV level diagram showing the states considered in our model, which includes the ground state triplet $m_s = 0$ state ($GS_0$), the ground state triplet $m_s = \pm1$ states ($GS_1$), the excited state triplet $m_s = 0$ state ($ES_0$), the excited state triplet $m_s = \pm1$ states ($ES_1$), the singlet state $A_1$ and the NV$^0$ ground ($GS_{NV^{0}}$) and excited ($ES_{NV^{0}}$) states, along with all corresponding transition rates. } 
	\label{fig:s_Fig2}
\end{figure}

In order to quantify the optically induced spin dynamics of our NV centers in the presence of NV ionization and recombination, we expand upon a model introduced in previous works \cite{Robledo2011, Hacquebard2018, Roberts2019, Manson2006}. A representation of the states of our system and the corresponding transition rates can be found in Fig.~\ref{fig:s_Fig2}. In our model, we use five levels to describe the NV$^-$ spin states, which include two states to describe the ground state triplet ($GS_0$ and $GS_1$), two states from the excited state triplet ($ES_0$ and $ES_1$), and a single state for the singlet manifold ($A_1$). We also include ground and excited states for NV$^0$ ($GS_{NV^0}$ and $ES_{NV^0}$). This allows us to describe the system as a state vector

\begin{equation}
    \B{\mathcal{\rho}_{\psi}(t)} = 
    \begin{bmatrix}
    \rho_{ES_{1}}(t) \\
    \rho_{ES_{0}}(t) \\
    \rho_{A_{1}}(t) \\
    \rho_{GS_{1}}(t) \\
    \rho_{GS_{0}}(t) \\
    \rho_{ES_{NV^{0}}}(t) \\
    \rho_{GS_{NV^{0}}}(t)
    \end{bmatrix},
\end{equation}

\noindent where the entries of the vector $\B{\mathcal{\rho}_{\psi}}$ describe the probability of finding the system in each of the seven states in our model. 

To model the dynamics of the system, we consider the set of transitions that couple all of these spin states, ignoring spin non-conserving transitions (aside from the intersystem crossing and ionization/recombination which is addressed later). This includes a transition rate $\Gamma_{532}$ between the ground triplet states and the excited triplet states, a transition rate $\Gamma_{ES}$ between the excited triplet states and the ground triplet states, and a transition rate $\Gamma_{ES_{0,1}\rightarrow A_1}$  between the triplet excited states and the singlet manifold. We also include a transition rate $\Gamma_{A_1}$ for the decay from the singlet state to the $GS_0$ and $GS_1$ states where the so-called ``branching ratio" $\mathcal{P}_{A_1\rightarrow GS_1}$ describes the probability of decay into $GS_1$ (with $1-\mathcal{P}_{A_1\rightarrow GS_1}$ describing the probability of decay into $GS_0$). Finally, we include the transition rate $\Gamma_{\text{ion}}$ from the NV$^-$ excited states ($ES_0$ and $ES_1$) to the NV$^0$ ground state ($GS_{NV^0}$) and the transition rate $\Gamma_{\text{rec}}$ from the NV$^0$ excited state ($ES_{NV^0}$) to NV$^-$ ground states ($GS_1$ and $GS_0$).
We assume that both $ES_0$ and $ES_1$ ionize at the same rate $\Gamma_{\text{ion}}$ and that recombination from the NV$^0$ excited state goes into the three spin states of NV$^-$ with equal probabilities \cite{Chen2015}. With these transition rates the dynamics of the system can be described by the following coupled rate equations:

\begin{widetext}
\begin{equation}
  \begin{aligned}
\dot{\rho}_{ES_1} &= -\left(\Gamma_{ES}+\Gamma_{ES_1\rightarrow A_1}+\Gamma_{\text{ion}}\right)  \rho_{ES_1}  + \Gamma_{532}  \rho_{GS_1} \\
\dot{\rho}_{ES_0} &= -\left(\Gamma_{ES}+\Gamma_{ES_0\rightarrow A_1}+\Gamma_{\text{ion}}\right)  \rho_{ES_0}  + \Gamma_{532}  \rho_{GS_0} \nonumber \\
\dot{\rho}_{A_1} &=  \Gamma_{ES_1 \rightarrow A_1} \rho_{ES_1} + \Gamma_{ES_0 \rightarrow A_1} \rho_{ES_0}  -\Gamma_{A_1}\rho_{A_1}\nonumber\\
\dot{\rho}_{GS_1} &=   \Gamma_{ES} \rho_{ES_1} +  \mathcal{P}_{A_1 \rightarrow GS_1}\Gamma_{A_1} \rho_{A_1} - \Gamma_{532} \rho_{GS_1} + (2\Gamma_{\text{rec}}/3)\rho_{ES_{NV^0}} \nonumber\\
\dot{\rho}_{GS_0} &=  \Gamma_{ES} \rho_{ES_0} +  (1-\mathcal{P}_{A_1 \rightarrow GS_1})\Gamma_{A_1} \rho_{A_1}  - \Gamma_{532} \rho_{GS_0}  + (\Gamma_{\text{rec}}/3)\rho_{ES_{NV^0}} \nonumber\\
\dot{\rho}_{ES_{NV^0}} &= - (\Gamma_{\text{rec}} + \Gamma_{ES_{NV^0}}) \rho_{ES_{NV^0}} + \Gamma_{532, NV^0} \rho_{GS_{NV^0}} \nonumber\\
\dot{\rho}_{GS_{NV^0}} &= \Gamma_{\text{ion}} \rho_{ES_1} + \Gamma_{\text{ion}} \rho_{ES_0} + \Gamma_{ES_{NV^0}} \rho_{ES_{NV^0}} - \Gamma_{532, NV^0} \rho_{GS_{NV^0}} \nonumber,
\end{aligned}
\end{equation}
\end{widetext}

\noindent which can also be expressed in a matrix form

\begin{equation}
    \B{\dot{\mathcal{\rho}}_{\psi}}(t) = \B{R_M} \B{\mathcal{\rho}_{\psi}}(t)
\end{equation}

\noindent
\begin{widetext}
\begin{small}
\begin{equation*}
    \B{R_M} = 
    \begin{bmatrix}
    -\left(\Gamma_{ES}+\Gamma_{ES_1\rightarrow A_1}+\Gamma_{\text{ion}}\right) & 0 & 0 & \Gamma_{532} & 0 & 0 & 0\\
     0 & -\left(\Gamma_{ES}+\Gamma_{ES_0\rightarrow A_1}+\Gamma_{\text{ion}}\right) & 0 & 0 & \Gamma_{532} & 0 & 0\\
     \Gamma_{ES_1 \rightarrow A_1} & \Gamma_{ES_0 \rightarrow A_1} & -\Gamma_{A_1} & 0 & 0 & 0 & 0\\
     \Gamma_{ES} & 0 & \mathcal{P}_{A_1 \rightarrow GS_1}\Gamma_{A_1} & -\Gamma_{532} & 0 & 2\Gamma_{\text{rec}}/3 & 0\\
     0 & \Gamma_{ES} & (1-\mathcal{P}_{A_1 \rightarrow GS_1})\Gamma_{A_1} & 0 & -\Gamma_{532} & \Gamma_{\text{rec}}/3 & 0\\
     0 & 0 & 0 & 0 & 0 & -\Gamma_{\text{rec}}-\Gamma_{ES_{NV^0}} & \Gamma_{532_{NV^0}} \\
     \Gamma_{\text{ion}} & \Gamma_{\text{ion}} & 0 & 0 & 0 & \Gamma_{ES_{NV^0}} & -\Gamma_{532_{NV^0}}
    \end{bmatrix}.
    \label{equ:rate_equ_mat}
\end{equation*}
\end{small}
\end{widetext}

For the readout, the NV center is initialized with the same green pulse. This means that the initial charge state can be estimated by solving for the steady state of the rate equation matrix ($\B{\dot{\mathcal{\rho}}_{\psi}}(t) = 0$). The solution can be found as the zero eigenvector of $\B{R_M}$, $\B{\mathcal{\rho}_{\psi,S.S.}}$. We then sum the components of $\B{\mathcal{\rho}_{\psi,S.S.}}$ from the NV$^0$ manifold and the NV$^-$ manifold, which we call $\mathcal{P}_{NV^0}$ and $\mathcal{P}_{NV^-}$, respectively. Right before the readout, we assume that the NV is in the ground state because we have waited for a time longer than the excited state lifetimes. This means that if we initialize in $m_s = 0$ then:
\begin{equation}
    \B{\mathcal{\rho}_{\psi}}(t=0) = 
    \begin{bmatrix}
    \rho_{ES_1}(t=0) \\
    \rho_{ES_0}(t=0) \\
    \rho_{A_1}(t=0) \\
    \rho_{GS_1}(t=0) \\
    \rho_{GS_0}(t=0) \\
    \rho_{ES_{NV^0}}(t=0) \\
    \rho_{GS_{NV^0}}(t=0) \\
    \end{bmatrix} =
    \begin{bmatrix}
    0 \\
    0 \\
    0 \\
    0 \\
    \mathcal{P}_{NV^-}  \\
    0 \\
    \mathcal{P}_{NV^0}
    \end{bmatrix},
\end{equation}
and if we initialize in $m_s = -1$ then:
\begin{equation}
    \B{\mathcal{\rho}_{\psi}}(t=0) = 
    \begin{bmatrix}
    \rho_{ES_1}(t=0) \\
    \rho_{ES_0}(t=0) \\
    \rho_{A_1}(t=0) \\
    \rho_{GS_1}(t=0) \\
    \rho_{GS_0}(t=0) \\
    \rho_{ES_{NV^0}}(t=0) \\
    \rho_{GS_{NV^0}}(t=0) \\
    \end{bmatrix} =
    \begin{bmatrix}
    0 \\
    0 \\
    0 \\
    \mathcal{P}_{NV^-} \\
    0  \\
    0 \\
    \mathcal{P}_{NV^0}
    \end{bmatrix}.
\end{equation}

These initial states are then evolved in time according to the rate equation matrix to give

\begin{equation}
    \B{\rho_{\psi}} (t)  = e^{\B{R_M} t} \B{\rho_{\psi}} (t=0). 
\end{equation}

The PL measured in the experiment is predominantly from transitions from the excited state triplet to the ground state triplet of NV$^-$(which occurs at a rate $\Gamma_{ES}$) or from the excited state to ground state of NV$^0$ (which occurs at a rate $\Gamma_{ES_{NV^0}}$). This means that given the model predictions for the population in each of the states as a function of time, we can calculate the PL as

\begin{equation}\label{equ:plt}
    PL(t) \propto \Gamma_{ES} \left( \rho_{ES_1}(t) + \rho_{ES_0}(t) \right)+  \Gamma_{ES_{NV^0}} \rho_{ES_{NV^0}}(t).
\end{equation}

To obtain the actual PL, we need to take into account the total collection efficiency of our measurement setup. This will introduce an overall scale factor to Eq.~\ref{equ:plt} and convert the proportionality ($\propto$) to an equality ($=$). However, in order to reduce the number of fit parameters in our model, we normalize both the experimental data and the predicted PL from the model.

\section{\romannumeral 4. Curve Fitting Procedure and Results}\label{subsec:curveFittingProcedure}

The PL traces of NV fluorescence were taken using the picoharp with a sampling resolution of 128 ps. Before fitting, the data are smoothed by averaging every 100 consecutive points. In order to remove the finite rise time of the AOM, we then search for the point after crossing our threshold where the instantaneous derivative of the PL changes from positive to negative, which we define as the $t=0$ point of the time trace. Each dataset is then normalized such that the steady-state PL = 1 (as described in the previous section). Then, we combine the data trace for the $m_s=0$ ($GS_0$) initialization and the $m_s=-1$ ($GS_1$) initialization for eight different laser powers $p_{532} \in [p_0,p_1,..,p_n]$, where $n=8$. We then fit the set of sixteen curves with a single model that outputs all sixteen curves for a given set of parameters as a concatenated array. Fitting all the data sets in this way allows us to get a more precise measure of the NV parameters without overfitting.

To further avoid overfitting, we use a fit parameter $\beta_{532}$ which is multiplied by our measured laser power, $p_{532}$, to get $\Gamma_{532}$.  Similarly, we expect the ionization and recombination rates to be roughly linear as a function of $p_{532}$. A linear scaling of $\Gamma_{\text{ion}}$ and $\Gamma_{\text{rec}}$ with $\Gamma_{532}$ yields the observed quadratic dependence of the charge state conversion rates as shown in main text Fig.~3. For NV centers in sample F there was a slight power dependence of the NV$^0$ population and it generally decreased by a few percent across the full power range we measured. To account for this power dependence, we include additional parameters $\beta_{\text{ion},2}$ and $\beta_{\text{rec},2}$, which was multiplied by $p_{532}$ to give a slight quadratic scaling of $\Gamma_{\text{ion}}$ and $\Gamma_{\text{rec}}$. The net result was the following:

\begin{eqnarray}
\Gamma_{532}(p_{532}) &=& \beta_{532} p_{532} \\ 
\Gamma_{\text{ion}}(p_{532}) &=& \beta_{\text{ion}} p_{532} + \beta_{\text{ion},2} p_{532}^2 \\ 
\Gamma_{\text{rec}}(p_{532}) &=& \beta_{\text{rec}} p_{532} + \beta_{\text{rec},2} p_{532}^2,
\end{eqnarray}
where we have set the fit constraints such that $\beta_{\text{ion},2} \leq 0$ and $\beta_{\text{rec},2} \geq 0$ so that the NV$^0$ population is either constant as a function of $p_{532}$ or tends toward lower NV$^0$ population as $p_{532}$ increases. To further reduce the number of free parameters in our model, we fix the excitation rate from the ground state to the excited state of NV$^0$ as $\Gamma_{532, \text{NV}^0} \equiv \Gamma_{532}/3$ as is done in Ref.~\cite{Roberts2019}. In the final model, the only free parameters are $\beta_{532}$, $\beta_{\text{ion}}$, $\beta_{\text{ion},2}$,
$\beta_{\text{rec}}$, $\beta_{\text{rec},2}$, $\Gamma_{ES}$, $\Gamma_{ES_{NV^0}}$, $\Gamma_{ES_1\rightarrow A_1}$,
$\Gamma_{ES_0\rightarrow A_1}$, $\Gamma_{A_1}$, and $\mathcal{P}_{A_1 \rightarrow GS_1}$. 

We then construct a simple cost function which computes the norm squared difference between the predicted and measured time-resolved PL and use this to quantify the quality of our fits. To avoid local minima in the fitting procedure, we run a full fit for a range of guesses of $\beta_{\text{ion}}$ and select the fit with the overall smallest cost function. This is done because it is believed that the cost function landscape is nonconvex and therefore is difficult to navigate by the gradient descent method. Before considering the fit parameters of our full model, we fit the data to a simpler model that does not include NV$^0$ as a comparison. The parameters extracted from the fits using the full model are shown in Table~\ref{tab:full_model_params} and fits using a model which does not included ionization/recombination are shown in Table~\ref{tab:noIonRec_model_params}. Results here suggest that adding ionization and recombination consistently improve the fit performance while producing intrinsic NV parameters that are consistent with the literature. The primary deviation is in the excited state lifetimes, which deviate from rates in Ref.~\cite{Roberts2019} by at most a factor of two.

\section{\romannumeral 5. Effect of Ionization and Recombination on Contrast and Signal-to-Noise Ratio}\label{subsec:effectOnContrast}

\begin{figure}[tb]
	\begin{center}
		\includegraphics[width=8.6 cm]{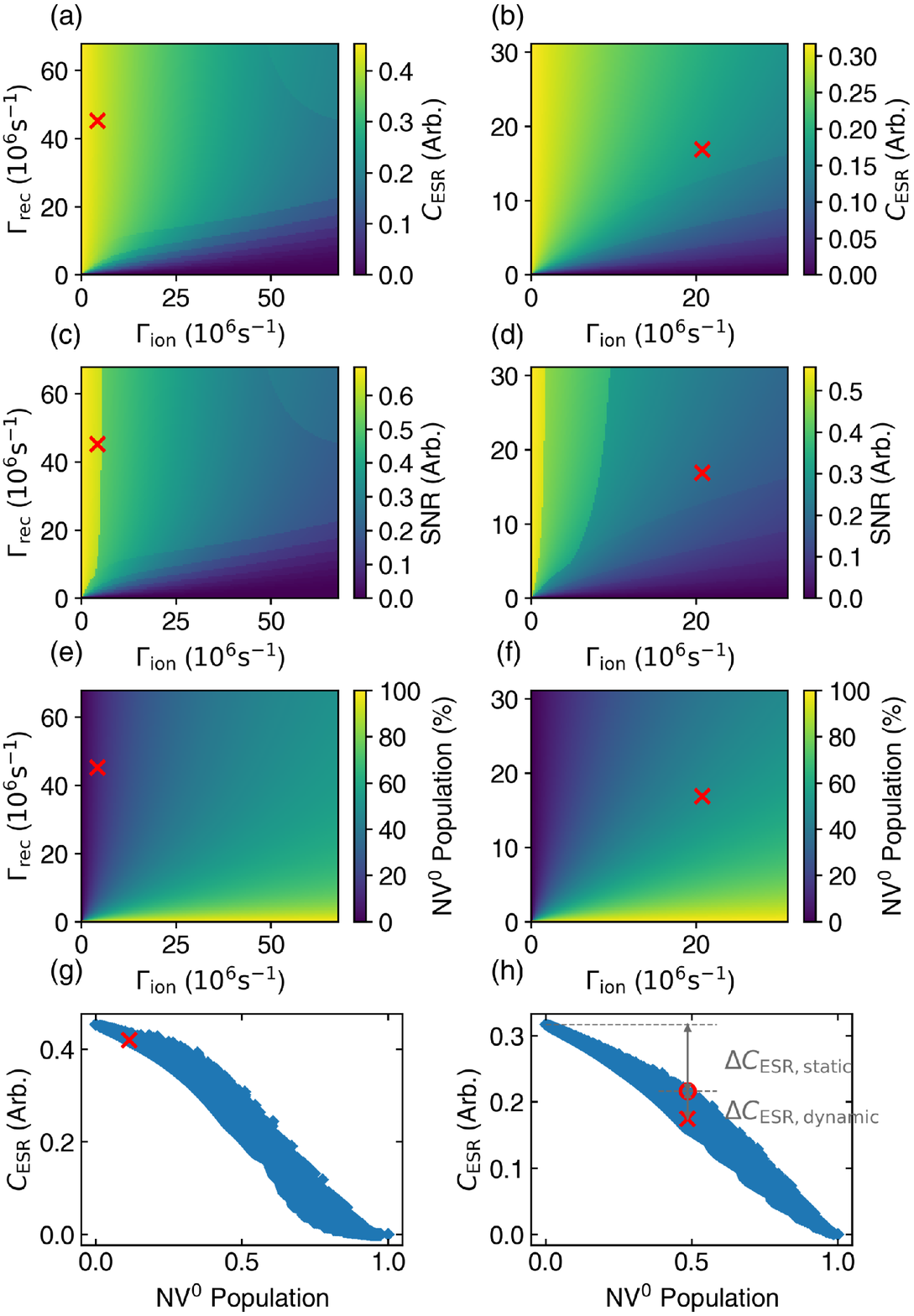}
	\end{center}
	\vspace{-0.6cm}
	\caption{(a), (b) Predicted OD-ESR contrast $C_{\text{ESR}}$ as a function of $\Gamma_{\text{ion}}$ and $\Gamma_{\text{rec}}$ using the extracted parameters from representative NV centers from main text Fig.~4 from sample A and sample F, respectively. (c), (d) Predicted SNR as a function of $\Gamma_{\text{ion}}$ and $\Gamma_{\text{rec}}$ for the same NV centers. (e), (f) Steady-state NV$^0$ population based on $\Gamma_{\text{ion}}$ and $\Gamma_{\text{rec}}$ for the same NV centers. (g), (h), Collected points from data in (a) and (e) or (b) and (f) that show $C_{\text{ESR}}$ as a function of the NV$^0$ population. The finite lower bound of $C_{\text{ESR}}$ is a result of sampling from a finite range of $\Gamma_{\text{ion}}$ and $\Gamma_{\text{rec}}$. Red crosses correspond to the fitted $\Gamma_{\text{ion}}$ and $\Gamma_{\text{rec}}$ for the two NV centers. Based on the experimental contrast and the extracted parameter values we calculate a change in a change in contrast due to finite NV$^0$ population, $\Delta C_{\text{ESR,static}}$, and due to the magnitude of the ionization and recombination rates $\Delta C_{\text{ESR,dynamic}}$.}
	\label{fig:s_Fig3}
\end{figure}

Given that our model fits the experimental data well, we can then see how $C_{\text{ESR}}$ depends on various parameters in the model. In particular, we are interested in how $C_{\text{ESR}}$ depends on $\Gamma_{\text{ion}}$ and $\Gamma_{\text{rec}}$. We begin by using the parameters which are extracted from fits of NV 17 in sample A and NV 91 in sample B. Then, we predict $C_{\text{ESR}}$ based on these parameters but instead of using parameters $\beta_{\text{ion}}$, $\beta_{\text{ion},2}$,
$\beta_{\text{rec}}$, $\beta_{\text{rec},2}$ we directly specify $\Gamma_{\text{ion}}$ and $\Gamma_{\text{rec}}$. Results of the expected $C_{\text{ESR}}$, which includes optimizing the readout window to maximize $C_{\text{ESR}}$, are plotted in Fig.~\ref{fig:s_Fig3}(a) and (b) for NV 17 in sample A and NV 91 in sample B, respectively. These are the same NV centers used in main text Fig.~4(b) and (c-e), respectively. The corresponding SNR following Ref.~\cite{Hopper2018} is plotted in Fig.~\ref{fig:s_Fig3}(c) and (d).  Red crosses correspond to the parameters from the fitted data which correspond to the power which maximizes $C_{\text{ESR}}$. In addition, based on $\Gamma_{\text{ion}}$ and $\Gamma_{\text{rec}}$ we calculate the NV$^0$ population which is shown in Fig.~\ref{fig:s_Fig3}(e) and (f). Finally, we use all of the $C_{\text{ESR}}$ values in Fig.~\ref{fig:s_Fig3}(a) and (b) and plot them against the NV$^0$ populations in Fig.~\ref{fig:s_Fig3}(e) and (f) to produce a scatter plot of $C_{\text{ESR}}$ [Fig.~\ref{fig:s_Fig3}(g, h)] as a function of NV$^0$ population, where it is clear that as the NV$^0$ population $\rightarrow$ 100\%,  $C_{\text{ESR}}$ $\rightarrow$ 0. The similar trend holds for SNR. In addition, using the values shown in Fig.~\ref{fig:s_Fig3}(g) and (h) we can quantify the decrease of contrast due to ionization and recombination by looking at the set of contrasts for a given NV$^0$ population, where we attribute a change in contrast due to the extracted NV$^0$, $\Delta C_{\text{ESR,static}}$, which is the difference between $C_{\text{ESR}}$ with all NV$^-$ population and the \emph{highest $C_{\text{ESR}}$ at that NV$^-$ population.} We can then attribute the change in contrast due to the magnitude of the ionization and recombination rates, $\Delta C_{\text{ESR,dynamic}}$, which is the difference between the $C_{\text{ESR}}$ of our NV center and the best $C_{\text{ESR}}$ at the same NV$^0$ population [see Fig.~\ref{fig:s_Fig3}(g) and (h) for a visual representation of this]. It should be noted that $C_{\text{ESR}}$ does not extend all the way to 0 because we have a finite upper bound for $\Gamma_{\text{ion}}$ and $\Gamma_{\text{rec}}$ in Fig.~\ref{fig:s_Fig3}(a-f). 

\section{\romannumeral 6. Relationship Between $\Gamma_{\text{ion},\text{rec}}$ and $r_{\text{ion},\text{rec}}$}\label{subsec:r-Gamma relation}

To elucidate the relationship between $\Gamma_{\text{ion},\text{rec}}$ and $r_{\text{ion},\text{rec}}$ we consider a 4-level system, which consists of the ground state and excited state for NV$^-$ ($GS_{NV^-}$: level 1 and $ES_{NV^-}$: level 2) and for NV$^0$ ($GS_{NV^0}$: level 3 and $ES_{NV^0}$: level 4). We use $\rho_i$ to represent the population in level $i$ and from the conservation of total population we have $\sum_{i=1}^4 \rho_i = 1$.

We assume the conversion rate from level $i$ to level $j$ is $\Gamma_{ij}$. For each NV charge state we have photon-induced excitation rates $\Gamma_{12}$, $\Gamma_{34}$ and emission rates $\Gamma_{21}$, $\Gamma_{43}$. Also, we include the ionization rate from the NV$^-$ excited state to the NV$^0$ ground state $\Gamma_{23}$ and the recombination rate from the NV$^0$ excited state to the NV$^-$ ground state $\Gamma_{41}$.

The resultant rate equations are:

\begin{equation}
  \begin{aligned}
  	&\dot{\rho_1} = -\Gamma_{12} \rho_1 + \Gamma_{21} \rho_2 + \Gamma_{41} \rho_4 \\
    &\dot{\rho_2} = \Gamma_{12} \rho_1 - \Gamma_{21} \rho_2 - \Gamma_{23} \rho_2 \\
    &\dot{\rho_3} = \Gamma_{23} \rho_2 - \Gamma_{34} \rho_3 + \Gamma_{43} \rho_4 \\
    &\dot{\rho_4} = \Gamma_{34} \rho_3 - \Gamma_{43} \rho_4 - \Gamma_{41} \rho_4
  \end{aligned},
 \nonumber
\end{equation} 

\noindent which can be rewritten as 

\begin{equation}
    \B{\dot{\mathcal{\rho}}_{\psi,4}}(t) = \B{R_{M,4}} \B{\mathcal{\rho}_{\psi,4}}(t),
\end{equation}

\noindent where 

\begin{equation}
    \B{R_{M,4}} = 
    \begin{pmatrix}
    -\Gamma_{12} & \Gamma_{21} & 0 & \Gamma_{41}\\
    \Gamma_{12}  & -\Gamma_{21}-\Gamma_{23} & 0 & 0\\
    0 & \Gamma_{23} & -\Gamma_{34} & \Gamma_{43}\\
    0 & 0 & \Gamma_{34} & -\Gamma_{43}-\Gamma_{41}
    \end{pmatrix}
\end{equation}

\noindent and $\B{\mathcal{\rho}_{\psi,4}} = (\rho_1, \rho_2, \rho_3, \rho_4)^T$.

Starting from this full picture, we want to consider the conversion rates between the two charge states: $\rho_{NV^-} \equiv \rho_{1} + \rho_{2}$ and $\rho_{NV^0} \equiv \rho_3 + \rho_4$. Thus, the time derivative of the two charge state populations will be 

\begin{eqnarray}
  	\dot{\rho}_{NV^-} &=& - \Gamma_{23} \rho_2 + \Gamma_{41} \rho_4 \nonumber \\
  	&=& - \Gamma_{23} \frac{\rho_2}{\rho_{NV^-}}\rho_{NV^-} + \Gamma_{41}  \frac{\rho_4}{\rho_{NV^0}}\rho_{NV^0}\\
    \dot{\rho}_{NV^0} &=& \Gamma_{23} \rho_2 - \Gamma_{41} \rho_4 \nonumber \\
    &=&  \Gamma_{23} \frac{\rho_2}{\rho_{NV^-}}\rho_{NV^-} - \Gamma_{41}  \frac{\rho_4}{\rho_{NV^0}}\rho_{NV^0} 
\end{eqnarray}

In general if the ratios $\rho_2/\rho_{NV^-}$ and $\rho_4/\rho_{NV^0}$ change with time, we will not observe the exponential decay or increase in the populations. However, if the charge state conversion rates are much slower than the internal transition rates for each charge state, i.e. \{$\Gamma_{23}$, $\Gamma_{41}$\} $\ll$ \{$\Gamma_{12}$, $\Gamma_{21}$, $\Gamma_{34}$, $\Gamma_{43}$\}, then in a timescale shorter than the charge state conversion timescale, the populations will reach a steady state within each charge state:

\begin{equation}
\left\{
  \begin{aligned}
  	&\dot{\rho_1} = -\Gamma_{12} \rho_1 + \Gamma_{21} \rho_2 = 0 \\
    &\dot{\rho_2} = \Gamma_{12} \rho_1 - \Gamma_{21} \rho_2 = 0 
  \end{aligned}
\right.  
\end{equation}

\begin{equation}
\left\{
  \begin{aligned}
    &\dot{\rho_3} = - \Gamma_{34} \rho_3 + \Gamma_{43} \rho_4 = 0 \\
    &\dot{\rho_4} = \Gamma_{34} \rho_3 - \Gamma_{43} \rho_4 = 0
  \end{aligned}
\right.  
\end{equation}

This leaves us with 
\begin{equation}
	\frac{\rho_2}{\rho_{NV^-}} = \frac{\Gamma_{12}}{\Gamma_{12}+\Gamma_{21}},\quad \frac{\rho_4}{\rho_{NV^0}} = \frac{\Gamma_{34}}{\Gamma_{34}+\Gamma_{43}},
\end{equation}

Meaning that we can write the rate equation for two charge states as 

\begin{equation}
\left\{
  \begin{aligned}
  	&\dot{\rho}_{NV^-} = - r_{\text{ion}}\rho_{NV^-} + r_{\text{rec}}\rho_{NV^0} \\
    &\dot{\rho}_{NV^0} =  r_{\text{ion}}\rho_{NV^-} - r_{\text{rec}}\rho_{NV^0}
  \end{aligned}
  ,
\right.  
\end{equation}
\noindent where $r_{\text{ion}}=\Gamma_{23} \Gamma_{12}/(\Gamma_{12}+\Gamma_{21})$ and $r_{\text{rec}} = \Gamma_{41} \Gamma_{34} / (\Gamma_{43} + \Gamma_{34})$.

Now that we have derived effective rate equations for the populations in each of the charge states, we can explore how each of the individual transition rates affects the overall transition rate between charge states. The excitation from the ground state to the excited state, ionization, and recombination are each one-photon processes, which means that the rates $\Gamma_{12}$, $\Gamma_{34}$, $\Gamma_{23}$, and $\Gamma_{41}$ should each have linear power dependence. $\Gamma_{21}$ and $\Gamma_{43}$, which are the spontaneous emission rates analogous to $\Gamma_{ES}$ and $\Gamma_{ES_{NV^0}}$, are decided by the intrinsic properties of NV centers and should not depend on power. If the laser power is much lower than the saturation power then $\Gamma_{12}\ll\Gamma_{21}$ and $\Gamma_{34}\ll\Gamma_{43}$. This means that $r_{\text{ion}}$ and $r_{\text{rec}}$ will each have a quadratic power dependence. If the laser power is much higher than the saturation power, then $\Gamma_{12}\gg\Gamma_{21}$ and $\Gamma_{34}\gg\Gamma_{43}$. This means that $r_{\text{ion}}$ and $r_{\text{rec}}$ should have linear power dependence, which agrees with the measurement results in Ref.~\cite{Waldherr2011}.
This analysis demonstrates an intuitive relationship between individual transition rates in our rate equation model and $r_{\text{ion}}$ and $r_{\text{rec}}$ in various parameter regimes. It also qualitatively explains the power dependence of $r_{\text{ion}}$ and $r_{\text{rec}}$ in the main text Fig.~3. It is worth noting, however, that the model here is a minimal model of the full rate equation model used in this paper but we believe the same intuition holds for the full model albeit with slightly more cumbersome expressions.

\begin{table*}[]
\begin{tiny}
\centering
\begin{ruledtabular}
\begin{tabular}{cccccccccccccc}
Sample & NV \# & $C_{\text{ESR}}$ &  $\mathcal{P}_{NV^0}$ (\%) & $\Gamma_{532}$ & $\Gamma_{\text{ion}}$ (MHz) & $\Gamma_{\text{rec}}$ (MHz) & $\Gamma_{\text{ES}}$ (MHz) & $\Gamma_{\text{ES} NV^0}$ (MHz) & $ES_0$ $\tau$ (ns) & $ES_1$ $\tau$ (ns) & $A_1$ $\tau$ (ns) & $\mathcal{P}_{A_1\rightarrow GS_1}$ & Cost     \\ \hline
A      & 10    & 0.21            & 0.11               & 19.6           & 5.6                        & 390.3                      & 75                        & 36                             & 10               & 8                & 60              & 0.25                     & 1.79E-05 \\
A      & 16    & 0.18            & 0.32               & 24.9           & 21.2                       & 176.6                      & 75                        & 36                             & 12               & 10               & 104             & 0.25                     & 3.34E-05 \\
A      & 17    & 0.38            & 0.11               & 24.2           & 4.2                        & 45.2                       & 75                        & 16                             & 12               & 8                & 104             & 0.25                     & 4.57E-05 \\
A      & 11    & 0.34            & 0.08               & 14.0           & 2.8                        & 74.8                       & 75                        & 16                             & 12               & 8                & 88              & 0.25                     & 8.53E-05 \\
F      & 78    & 0.15            & 0.30               & 18.3           & 5.5                        & 8.3                        & 75                        & 36                             & 8                & 6                & 62              & 0.37                     & 9.85E-06 \\
F      & 91    & 0.16            & 0.49               & 22.4           & 20.7                       & 16.9                       & 70                        & 32                             & 11               & 8                & 124             & 0.25                     & 1.02E-05 \\
F      & 40    & 0.21            & 0.21               & 11.1           & 0.1                        & 0.2                        & 75                        & 20                             & 9                & 7                & 69              & 0.25                     & 2.96E-05 \\
F      & 24    & 0.27            & 0.25               & 9.4            & 3.8                        & 6.3                        & 75                        & 16                             & 10               & 7                & 86              & 0.25                     & 3.97E-05
\end{tabular}
\end{ruledtabular}
\caption{Fitted parameters based on the rate equation model incorporating ionization and recombination. Listed $C_{\text{ESR}}$, $\mathcal{P}_{NV^0}$, $\Gamma_{532}$, $\Gamma_{\text{ion}}$, and $\Gamma_{\text{rec}}$ are the values at the maximum $C_{\text{ESR}}$ across $p_{532}$ (and resultant $\Gamma_{532}$). We also include effective lifetimes $ES_0$ $\tau$ = 1/($\Gamma_{ES}$ + $\Gamma_{ES_0\rightarrow A_1}$), $ES_1$ $\tau$ = 1/($\Gamma_{ES}$ + $\Gamma_{ES_1\rightarrow A_1}$), and $A_1$ $\tau$ = 1/$\Gamma_{A_1}$. It is worth noting that the actual excited state lifetime should also include ionization and recombination but we have selected these definitions to compare to models without ionization and recombination. $\Gamma_{ES}$ is constrained to be $<75$ MHz and we are saturating the bound in our fit procedure.}
\label{tab:full_model_params}
\end{tiny}
\end{table*}

\begin{table*}[]
\begin{tiny}
\centering
\begin{ruledtabular}
\begin{tabular}{cccccccccc}
Sample & NV \# & $C_{\text{ESR}}$ & $\Gamma_{532}$ (MHz) & $\Gamma_{\text{ES}}$ (MHz) & $ES_0$ $\tau$ (ns) & $ES_1$ $\tau$ (ns) & $A_1$ $\tau$ (ns) & $\mathcal{P}_{A_1\rightarrow GS_1}$ & Cost     \\ \hline
A      & 10    & 0.21            & 16.1           & 68                        & 10               & 8                & 56              & 0.25                     & 2.18E-05 \\
A      & 17    & 0.38            & 20.5           & 68                        & 12               & 8                & 93              & 0.25                     & 7.71E-05 \\
A      & 16    & 0.18            & 11.8           & 45                        & 11               & 9                & 63              & 0.35                     & 8.09E-05 \\
A      & 11    & 0.34            & 11.5           & 67                        & 12               & 8                & 81              & 0.25                     & 1.20E-04 \\
F      & 78    & 0.15            & 13.6           & 44                        & 11               & 10               & 49              & 0.40                     & 2.47E-05 \\
F      & 91    & 0.16            & 9.4            & 33                        & 11               & 9                & 62              & 0.30                     & 4.29E-05 \\
F      & 40    & 0.21            & 11.3           & 63                        & 11               & 9                & 53              & 0.25                     & 4.57E-05 \\
F      & 24    & 0.27            & 8.7            & 49                        & 13               & 9                & 66              & 0.25                     & 9.44E-05
\end{tabular}
\end{ruledtabular}
\caption{Fitted parameters based on the rate equation model without ionization and recombination.}
\label{tab:noIonRec_model_params}
\end{tiny}
\end{table*}

\bibliographystyle{apsrev4-1}
\bibliography{chargeState.bib}